\documentclass[
  reprint,        
  amsmath,amssymb,
  aps,prb,        
  floatfix,
  superscriptaddress
]{revtex4-2}

\usepackage[british]{babel}
\usepackage{graphicx}
\usepackage{xcolor}
\usepackage{bm}
\usepackage{siunitx}
\usepackage{xr-hyper}
\usepackage[colorlinks=true,allcolors=blue]{hyperref}

\newcommand{\software}[1]{\texttt{#1}}

\begin{document}

\title{A GPU-Native Nested Sampling Framework for Finite-Temperature
       Materials Modelling with Machine-Learned Interatomic Potentials}

\author{N.~Unglert}
\affiliation{Institute of Materials Chemistry, TU Wien, 1060 Vienna, Austria}
\email[Correspondence email address: ]{nico.unglert@tuwien.ac.at}
\author{A.~Serafini}
\affiliation{Institute of Materials Chemistry, TU Wien, 1060 Vienna, Austria}
\author{J. Carrete}
\affiliation{Instituto de Nanociencia y Materiales de Aragón (INMA), CSIC-Universidad de Zaragoza, 50009 Zaragoza, Spain}
\author{G. K. H. Madsen}
\affiliation{Institute of Materials Chemistry, TU Wien, 1060 Vienna, Austria}

\date{\today}

\begin{abstract}
Modern equivariant machine-learned interatomic potentials (MLIPs) have
dramatically improved the accuracy of and the scales accessible to  atomistic
simulation, yet the thermodynamic sampling methods needed to turn those
potentials into finite-temperature phase stability predictions have not
kept pace computationally. We present the GPU-native, massively parallel
nested sampling framework \texttt{JAXRENS}, implemented on top of \software{JAX}, which removes the practical bottlenecks preventing atomistic nested sampling from being combined with
state-of-the-art MLIPs. The framework provides distributed multi-GPU
execution, vectorised ensembles of walkers, and direct interoperability
with modern force fields.
We demonstrate the highly beneficial scaling of \texttt{JAXRENS} on hardware accelerators, and apply it to the simulation of the ZnS phase diagram based on a \software{MACE} foundation model, a procedure previously computationally inaccessible. 
Despite requiring in the order of $10^{10}$ model evaluations, the final converged simulation completed within six days on four GPUs, marking an important step towards routine, first-principles-quality predictions of finite-temperature phase stability using machine-learned interatomic potentials.
\end{abstract}

\maketitle

\section{Introduction}
\label{sec:intro}

Over the past several years, machine-learned interatomic potentials (MLIPs), and in particular equivariant message-passing architectures \cite{batatia2022mace,batzner2022nequip}, have transformed atomistic simulation by delivering near first-principles accuracy at a small fraction of the cost.
More recently, the field has moved towards foundation models: potentials pre-trained on vast, chemically diverse datasets that aim to describe much of the structural and chemical spaces out of the box and offer the possibility to be fine-tuned for specific systems \cite{batatia2025foundation,nequix}.

The thermodynamic sampling needed to turn state-of-the-art MLIPs into finite-temperature predictions, however, remains a computational bottleneck.
Converging a single free energy requires a very large number of energy and force evaluations, and a phase diagram multiplies that cost across many temperatures, pressures, and compositions.
Over the past decade, nested sampling (NS), originally introduced for Bayesian evidence computation\cite{skilling2006}, has emerged as a powerful tool for addressing this challenge \cite{partay2010,baldock2016}.
By directly evaluating the partition function, nested sampling can provide the free energy and other thermodynamic observables across an entire temperature range from a single run.
Crucially for materials applications, it captures phase transitions without predefined order parameters or prior knowledge of the competing phases.
These strengths have enabled applications ranging from pressure--temperature phase diagrams of bulk solids \cite{baldock2016,baldock_constant-pressure_2017,marchant2023carbon,unglert2023silicon,fletcher2025,hellyar2026} to surface phase behaviour and molecular clusters \cite{freebird,Riccius2026}.

NS works by maintaining a population of walkers that collectively map out the configuration space under increasingly restrictive energy or enthalpy constraints. 
More recently, replica-exchange nested sampling (RENS) \cite{unglert_replica_2025} has improved sampling efficiency and convergence by allowing configurations to be exchanged between simultaneously running simulations. This has substantially extended the reach of the method to larger and more complex systems and enabled active learning of MLIPs for first-principles-quality phase diagrams \cite{unglert_active_2026,fletcher2025,hellyar2026}.

The independent evolutions of the NS walkers are inherently parallel.
However, existing NS implementations are difficult to combine with GPU-native workflows, modern MLIPs, distributed hardware, and large atomistic systems.
Established atomistic NS codes, like \software{pymatnest} \cite{pymatnest} and the more recent \software{FreeBird.jl} \cite{freebird}, are CPU-oriented and were not designed to exploit the massively parallel throughput of modern accelerators. Meanwhile, differentiable, GPU-native engines such as \software{JAX-MD} \cite{jaxmd2020}, and more recently \software{torch-sim} \cite{torchsim2025} and \software{kUPS} \cite{kups2026}, have demonstrated the value of running many independent systems as batched, vectorised computations for atomistic simulations.
While these efforts mainly target molecular dynamics, Monte Carlo, and structural relaxation, they signal a broader shift towards batched, accelerator-native sampling. 

The importance of efficient batching is particularly pronounced in NS.
While the large number of required potential evaluations limits accessible system sizes, the computational cost of state-of-the-art MLIPs for these relatively small systems is often dominated by overhead rather than arithmetic workload.
Adopting a batched computational paradigm therefore offers a way to combine the broad thermodynamic sampling capabilities of NS with the throughput of modern GPUs.

Here, we bring this shift towards batched, accelerator-native computation to finite-temperature NS with the introduction of \software{JAXRENS}, a framework built on top of \software{JAX} \cite{bradbury_jax_2018} and designed around accelerator hardware and MLIPs from the outset.
Its guiding design choice is to treat the walker population itself as a batched, vectorised computation: the walkers of an iteration are advanced by a single compiled kernel, and whole simulations are distributed across one or more GPUs, which turns RENS from an add-on into a native mode of operation.
The sampler couples directly to modern MLIPs, among them \software{MACE}, \software{NeuralIL}, and \software{Nequix}, so that state-of-the-art models can drive the sampling without a per-model reimplementation.
In Sec.~\ref{sec:overview}, we outline the core functionality of \software{JAXRENS}, followed by numerical results in Sec.~\ref{sec:results} demonstrating the achieved speedups and their practical impact through a foundation-model-based study of the ZnS phase diagram, previously computationally prohibitive with existing codes.

\begin{figure*}[htbp]
  \centering
  \includegraphics[width=0.95\textwidth]{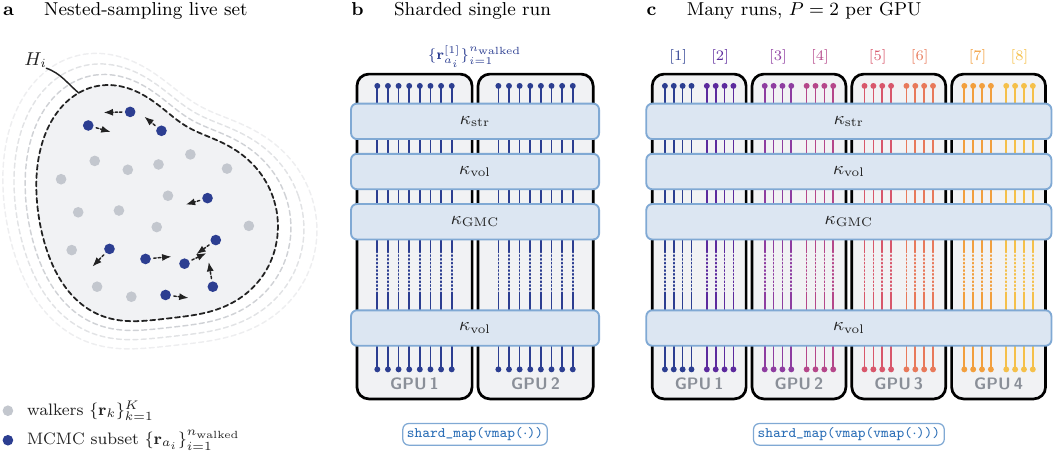}
  \caption{
  \textbf{Vectorised, multi-GPU execution of nested sampling.}
  (a)~A nested-sampling run maintains a live set of $K$ walkers; at each
  iteration the lowest-likelihood walkers are replaced and a subset of
  $n_\mathrm{walked}$ walkers is decorrelated by a sequence of constrained
  Monte Carlo move kernels (arrows), here cell-stretch
  ($\kappa_\mathrm{str}$), volume ($\kappa_\mathrm{vol}$), and Galilean
  Monte Carlo ($\kappa_\mathrm{GMC}$).
  (b)~A single run is sharded across several GPUs: its $n_\mathrm{walked}$ parallel
  walkers are split across devices and every move kernel is applied to the whole
  batch through \software{shard\_map(vmap($\cdot$))}.
  (c)~Many independent runs (superscripts $[1],[2],\dots$) are instead
  distributed across the GPUs, two per device here, each carrying its own walker
  batch, through \software{shard\_map(vmap(vmap($\cdot$)))}; this is the layout in which
  replica exchange couples the concurrent runs.
  }
  \label{fig:architecture}
\end{figure*}

\section{Framework Overview}
\label{sec:overview}

\subsection{The replica-exchange nested sampling algorithm}
\label{frame:ns}

NS explores configuration space through a population of $K$ uniformly distributed live points, constrained by an upper enthalpy bound. 
At each iteration, the live point with the highest enthalpy is removed and replaced by a new configuration sampled from the remaining constrained region via Markov-chain Monte Carlo (MCMC) of a cloned walker. 
As the enthalpy bound decreases, the accessible configuration space contracts, progressively resolving lower-enthalpy states. 
The resulting sequence of enthalpy levels and associated phase-space volumes allows thermodynamic properties to be
reconstructed over a continuous range of temperatures (see also Sec.~\ref{sec:meth_ns}). 
Different methods have been proposed for the parallelisation of the MCMC procedure, like removing multiple walkers per iteration \cite{martiniani_superposition_2014} or performing extra walks on a subset of additional walkers \cite{baldock_constant-pressure_2017}. 
In both cases, the workload boils down to the propagation of a set of $n_\mathrm{walked}$ independent Markov-chains (see Fig.~\ref{fig:architecture}a), an embarrassingly parallel problem.

RENS additionally couples $M$ individual NS simulations through occasional swaps between walker pools of neighbouring replicas \cite{unglert_replica_2025}. 
Since each swap involves only a pair of replicas, swaps between disjoint replica pairs can be performed in parallel.
Mixing along the entire replica axis is achieved by alternating between the parallel swap kernels $\mathcal{S}_\mathrm{even}$ and $\mathcal{S}_\mathrm{odd}$ (see Sec.~\ref{sec:meth_rens} for more detail). 
While this introduces additional implementation complexity (see Sec.~\ref{frame:replica_exchange}), the exchanges take place between the parallel MCMC decorrelation phases and are therefore largely orthogonal to the existing parallelisation schemes.

\subsection{General architecture and parallelisation}
\label{frame:architecture}

The \software{JAXRENS} framework expresses the entire nested sampling computation in terms of composable \software{JAX} transformations \cite{bradbury_jax_2018}, so that it executes natively on GPUs without bespoke kernels.
A single run maintains a live set of $K$ walkers and, at each iteration, decorrelates a subset of $n_\mathrm{walked}$ of them by applying a sequence of constrained Monte Carlo move kernels in lockstep.
Because every walker is advanced by the same kernel, the live set is stored as an array whose leading axis enumerates the $K$ walkers. At each iteration, a subset of $n_\mathrm{walked}$ walkers is selected, and a single just-in-time (JIT) compiled step kernel, targeting the constrained uniform prior at the current enthalpy limit $H_i$, is applied to this subset in parallel using \software{JAX}'s built-in vectorisation transformation \software{vmap}. A lightweight Python loop handles only the outer nested sampling bookkeeping and termination.
Adding a hardware axis on top of this walker axis is what distributes the work across $G$ GPUs: the \software{JAX} transformation \software{shard\_map} partitions a leading device axis of length $G$ over a one-dimensional mesh of devices, and \software{vmap} maps over the walker axis on each device, so the names of these transformations correspond directly to the axes of the walker array.
Because \software{shard\_map} composes with JIT compilation, the entire batched step, including its collective operations, is compiled into a single optimised program per device, with all inter-device communication handled automatically by Nvidia's \software{NCCL} library \cite{hu_demystifying_2026}.

This gives two complementary ways to use multiple GPUs (Fig.~\ref{fig:architecture}b,c).
On the one hand, a single large run can be \emph{sharded} across devices: its $K$ walkers are divided into $G$ groups of size $K/G$, forming a $G \times (K/G)$ array in which each GPU holds and advances $K/G$ walkers (\software{shard\_map(vmap($\cdot$))}), so that one simulation can exploit more memory and throughput than a single device provides.
On the other hand, $P$ independent runs can be placed on each device, forming a $G \times P \times K$ array of walkers --- $G$ GPUs, $P$ runs per GPU, $K$ walkers per run --- advanced by nesting a second \software{vmap} over the runs (\software{shard\_map(vmap(vmap($\cdot$)))}).
This second layout is what makes large ensembles of simulations, and the replica exchange that couples them, practical: the total workload remains substantial, but it is spread evenly across all available devices, so that calculations which would be prohibitively slow on a single GPU become feasible.

Two interfaces keep the framework general.
Energies and forces are supplied through a pluggable \software{EnergyBackend}, with shipped implementations for machine-learned potentials including \software{MACE}, \software{NeuralIL}, and \software{Nequix} that can be swapped in without touching the sampling code.
Since any \software{JAX}-differentiable potential is admissible, this set is not closed: PyTorch-to-\software{JAX} translation tools such as \software{tojax} \cite{tojax2026} bring torch-only models, such as foundation models of the FairChem/UMA family \cite{wood_uma_2026}, into reach without a per-model reimplementation, in principle opening the framework to much of the modern MLIP ecosystem.
The live set, in turn, is evolved by a configurable set of Monte Carlo moves built on a common \software{MoveKernel} interface.
The kernels shipped with the code cover random-walk, Galilean, and Hamiltonian Monte Carlo updates of the atomic positions, together with volume, stretch, and shear moves of the simulation cell, and their step sizes are tuned automatically during the run.
The command-line interface follows the same modular design, exposing both sets of components declaratively so that a simulation is specified in configuration rather than in code.
Altogether, \software{JAXRENS} is agnostic to the \software{JAX} energy backend, is readily extended by custom Monte Carlo moves -- for instance ones enforcing additional constraints \cite{yang_surface_2024} -- and provides a natural basis for implementing related population-based Monte Carlo methods such as sequential Monte Carlo \cite{salomone_unbiased_2025}.

\subsection{Replica exchange}
\label{frame:replica_exchange}

Having established the multi-run parallelization in Fig.~\ref{fig:architecture}c, the coupling of independent NS runs, becomes a natural extension.
Following the RENS scheme of Ref.~\cite{unglert_replica_2025} in the $NPT$ case (see also Sec.~\ref{sec:meth_rens}), a \emph{swap} attempts to exchange a single walker between two neighbouring simulations, and a \emph{swap round} --- triggered at a fixed interval of nested-sampling iterations --- sweeps over all non-overlapping neighbouring pairs so that every adjacent pair is offered an exchange. 
A swap is accepted whenever each exchanged configuration still satisfies the receiving simulation's nested-sampling likelihood constraint; because the stored enthalpy $U+PV$ is simply re-referenced to the partner's pressure, this test reuses quantities already in memory and never triggers a new evaluation of the potential.

\begin{figure}[htbp]
  \centering
  \includegraphics[width=\columnwidth]{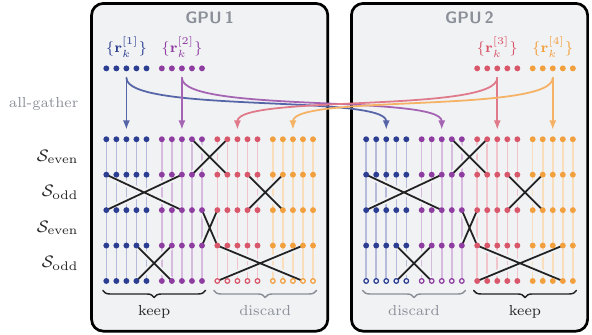}
  \caption{
  \textbf{Multi-GPU execution of a replica-exchange round.}
  Each replica $m$, from a total of $M=4$, carries its own live set $\{\mathbf{r}^{[m]}_k\}_{k=1}^{K}$ (columns, one colour per simulation), of which each GPU owns only a subset. 
  A single collective operation gathers the full population onto every device (curved arrows), so that each GPU holds a redundant copy of all $M$ live sets. 
  Every device then executes the identical sequence of even and odd swap kernels $\mathcal{S}_\mathrm{even}$, $\mathcal{S}_\mathrm{odd}$ (here $n_\mathrm{RE}=2$ rounds) from a shared random seed, so the swaps -- drawn as lines connecting neighbouring simulations -- are reproduced consistently without further communication.
  Afterwards, each device discards the copies it does not own and retains only its own updated live sets, so that inter-device traffic amounts to one gather per round rather than one exchange per attempted swap.
  }
  \label{fig:replica_exchange}
\end{figure}

Mechanically, the coupling acts along the run dimension that is already vectorised on the GPU. 
On a single device, a swap is nothing more than a pair of masked writes into the population arrays, so a full swap round adds negligible overhead. 
On multiple devices, where the runs are distributed across GPUs, Fig.~\ref{fig:replica_exchange}), a single collective operation per round gathers the full population onto every device; each device then reproduces the identical set of swaps from a shared random seed and retains only its own share, so inter-device communication is limited to one exchange of data per round rather than one per pair. 
In both cases the cost is dominated by the sampling itself, making large ensembles of coupled simulations a natural mode of operation.

\section{Results and discussion}
\label{sec:results}

\subsection{Scalability across MLIPs}
\label{sec:scaling}

\begin{figure}[t]
  \includegraphics*[width=\columnwidth]{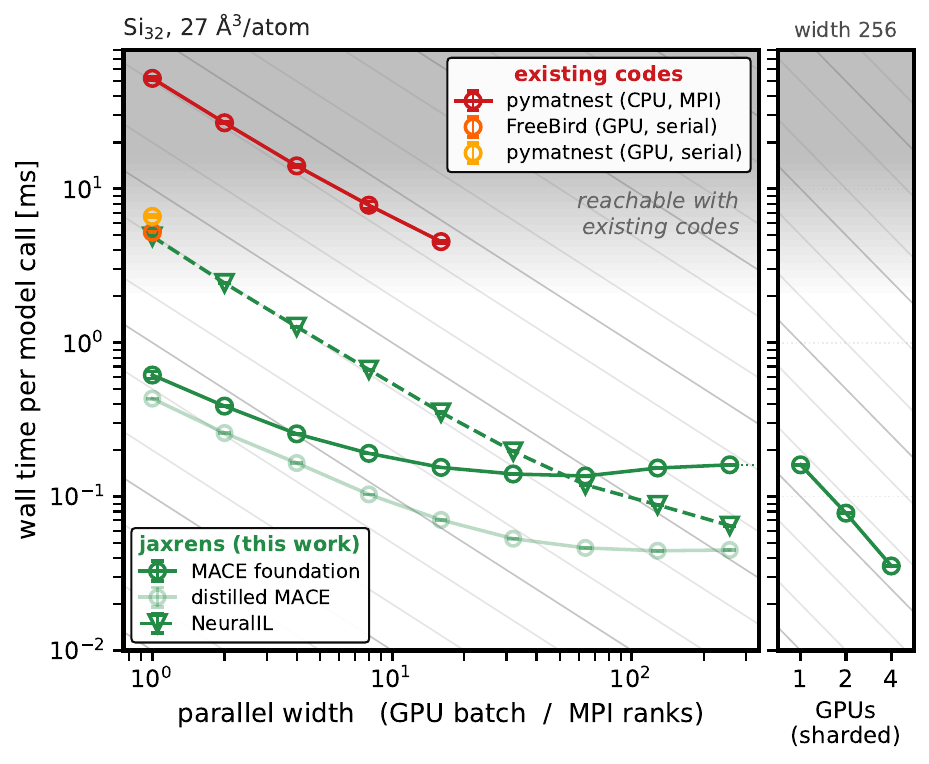}
  \caption{
  \textbf{Computational scalability across machine-learned potentials
  and nested-sampling drivers.}
  Wall time \emph{per model call} (one energy or energy-and-force evaluation of one configuration, counted in every driver) of a nested-sampling step for Si$_{32}$ at 27\,\AA$^3$/atom, versus the parallel width: the number of walkers evaluated as one GPU batch in our framework (green), the number of MPI ranks for \software{pymatnest} on the CPU (red line).
  \software{FreeBird.jl} and \software{pymatnest} on the GPU evaluate one configuration at a time and appear as single points at width~1.
  Results are shown for the \software{MACE-MP-0} foundation model~\cite{batatia2025foundation,deng_chgnet_2023}, a distilled compact \software{MACE} surrogate (Sec.~\ref{sec:meth_timing}) and a \software{NeuralIL} model~\cite{neuralil}.
  The grey shading marks the per-call times reachable with the existing codes; the thin diagonal lines are lines of perfect scaling (time $\propto 1/\text{width}$).
  Side panel: the foundation model at width~256 with the walkers sharded over 1, 2 and 4 GPUs (dotted: ideal $1/G$).
  Mean and standard deviation over five repeats with different random seeds; all GPU data on NVIDIA H100.
  }
  \label{fig:scaling}
\end{figure}

Figure~\ref{fig:scaling} reports the wall time per model call of a single nested-sampling step as a function of the intra-step parallelism, following the protocol of Sec.~\ref{sec:meth_timing}.
The parallelism nested sampling offers is the one \emph{across} walkers: the decorrelation of each cloned walker is a sequential Markov chain, but the walkers of an iteration are independent of one another.
How well a driver exploits that axis, and on what hardware, is what the comparison below isolates.

The established way of exploiting it is the one taken by \software{pymatnest} on the CPU, which distributes the per-walker work of a step over MPI ranks --- the same axis our framework batches over on the GPU.
This parallelisation is effective: going from one to sixteen ranks lowers the cost from 52 to 4.5~ms per model call, a speed-up of 11.5 (72\,\% parallel efficiency; red line in Fig.~\ref{fig:scaling}), so the walker axis itself poses no obstacle.
What limits this route is the hardware behind each rank.
Every rank evaluates the MLIP on a single CPU core, so even sixteen ranks only reach the cost of a serial GPU evaluation, and closing the further gap of one to two orders of magnitude to the batched GPU path would take far more cores than a node provides.

Moving the potential onto the GPU is the obvious remedy, and both \software{pymatnest} and \software{FreeBird.jl} can do so through their ASE calculator interface --- but only one configuration at a time.
In this mode the walker parallelism that MPI provided is given up, which is why these drivers appear as single points rather than as curves in Fig.~\ref{fig:scaling}.
The change of hardware alone brings the cost down to 5.2~ms (\software{FreeBird.jl}) and 6.6~ms (\software{pymatnest}) per model call, yet this falls far short of what the device can deliver.
Profiling a single \software{mace-torch} call through the ASE calculator (see SI) reveals that the GPU is active for only $\approx\!16$\,\% of the execution time, with most of the cost arising from Python overhead, graph construction, and kernel launches rather than network evaluation. Capturing the kernels in a CUDA graph reduces the bare model call from 9.5 to 1.8~ms, highlighting the inefficiency of evaluating one configuration at a time at this scale.

Our framework (solid green line in Fig.~\ref{fig:scaling}) recovers both at once, evaluating the $n_\mathrm{walked}$ walkers of a step as a single batched, just-in-time-compiled kernel.
Even without batching (width~1) it costs 0.62~ms per model call, $8$--$11\times$ less than the serial GPU drivers, although it evaluates energy \emph{and} forces in every call while those drivers mostly request the energy alone.
Part of this improvement stems from the MLIP implementation itself: XLA compilation makes \software{mace-jax} substantially faster than \software{mace-torch} even without batching (see SI). 
The larger gain, however, comes from amortising the remaining overhead across walkers. 
At a batch width of 64, the per-configuration time drops to 0.14~ms, more than $30\times$ faster than \software{pymatnest} on sixteen cores and $\approx\!40\times$ faster than the fastest serial GPU driver. 
By mapping walker-level parallelism directly onto the GPU, our framework exploits the accelerator efficiently without transferring control back to the host between evaluations.

Where saturation sets in --- and hence how much of this benefit is realised --- depends on the cost of the potential.
The \software{MACE-MP-0} foundation model saturates the GPU at a width of $\approx\!32$--$64$; beyond it the per-call time no longer falls and rises slightly, to 0.16~ms at a width of~256.
A cheaper potential pushes the floor down.
To demonstrate this, we distilled a compact \software{MACE} surrogate from the foundation model (Sec.~\ref{sec:meth_timing}) and also benchmark a descriptor-based \software{NeuralIL} model~\cite{neuralil}.
The distilled model saturates at a similar width but at a floor of 0.044~ms per call, $3\times$ below the foundation model.
\software{NeuralIL} shows the opposite regime: a single evaluation is expensive (4.9~ms at width~1), but it is almost entirely fixed cost, so that one batched evaluation of 256 configurations takes only $3.4\times$ as long as that of a single one.
Its per-call time therefore drops 75-fold to 0.065~ms at a width of~256 without saturating, and is below the foundation model from a width of~64 on.

Once a single GPU saturates, further speed-up requires more devices.
Sharding the walkers of one nested-sampling run over several GPUs (side panel of Fig.~\ref{fig:scaling}) reduces the per-call time of the foundation model at a width of~256 from 0.161~ms on one H100 to 0.078 and 0.035~ms on two and four, a factor of 4.5.
The speed-up exceeds the device count because each of the four GPUs evaluates a batch of 64 walkers, the width at which a single GPU runs most efficiently, rather than the 256 that a single GPU handles less well; measured against the best single-GPU point (0.136~ms at a width of~64), four GPUs are $3.9\times$ faster.
At 0.035~ms per call, the sharded run is $\approx\!130\times$ faster than \software{pymatnest} on sixteen CPU cores and $\approx\!150\times$ faster than the fastest serial GPU driver.

\begin{figure}[htbp]
  \includegraphics*[width=\columnwidth]{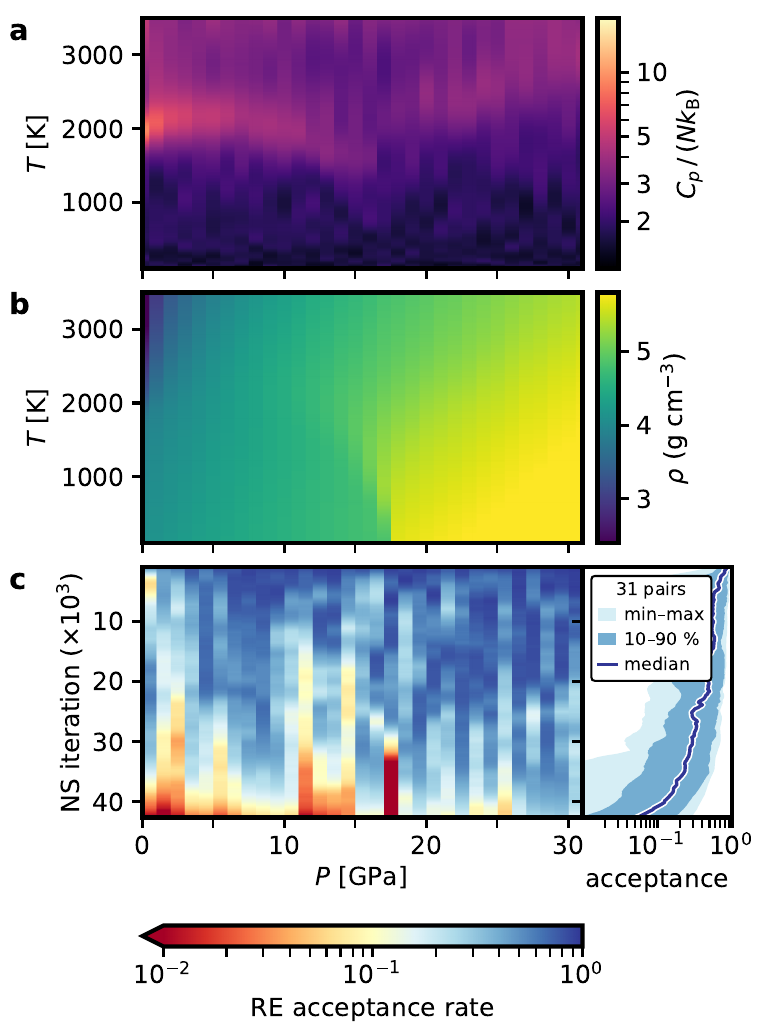}
  \caption{
    \textbf{Observables and diagnostics of a foundation model driven RENS simulation.}
    (a)~Heat capacity $C_p/(N k_\mathrm{B})$ and (b)~density $\rho$ from the NS partition function.
    (c)~Acceptance rate of the configuration swaps between neighbouring pressures, per pair of replicas and NS iteration (moving average over 2000 iterations; rates below 0.01 in the darkest red).
    The side panel shows the median, the 10--90\,\% range and the full range over the 31 pairs.
    NS iterations increase downwards, i.e.\ towards lower temperature.}
  \label{fig:re_acc}
\end{figure}

\subsection{Pressure--temperature phase diagram of ZnS}
\label{sec:zns}

Zinc sulfide is a prototypical tetrahedral semiconductor whose
pressure--temperature phase diagram is only partly explored by experiment. 
The pressure-induced transition from the fourfold-coordinated
zincblende (B3) to the sixfold-coordinated rocksalt (B1) structure has been
measured repeatedly, with room-temperature transition pressures scattering
between 12 and \SI{25}{GPa}, which has been attributed to differing pressure
calibrations and to the transition kinetics~\cite{ono2018}. In situ X-ray
diffraction above \SI{500}{K}, where hysteresis becomes negligible, yields
a negative Clapeyron slope, $P(\mathrm{GPa}) = 14.4 - 0.0033\,T(\mathrm{K})$
\cite{ono2018}, while low-temperature Raman measurements give transition
pressures rising to \SI{18.8}{GPa} at \SI{90}{K}~\cite{wu2025}. The melting
line, in contrast, is essentially unknown: ZnS sublimes at ambient pressure
and can only be melted under an inert-gas overpressure. Even the
ambient-pressure melting point is disputed, with \SI{1830 \pm 20}{\celsius}
(\SI{2103}{K}) measured under \SI{150}{psi} argon~\cite{addamiano1957} and
\SI{1700 \pm 5}{\celsius} (\SI{1973}{K}) for stoichiometric crystals, the
higher value being assigned to a Zn-rich composition~\cite{vasilyeva1998}.
To our knowledge, no measured or computed melting curve of ZnS under
pressure has been reported. Here we use RENS \cite{unglert_replica_2025} to compute the full $P$--$T$ phase diagram, including melting, directly from MLIPs.

\subsubsection{Foundation model phase diagram}

We start from the r\textsuperscript{2}SCAN variant of the MACE-MATPES-0 foundation model~\cite{batatia2022mace,kaplan2025matpes}, trained on the r\textsuperscript{2}SCAN~\cite{furness_accurate_2020} part of the MatPES dataset~\cite{kaplan2025matpes}, and use it as is, without any system-specific training.
With it we run $NPT$ nested sampling for a 16-atom cell at 32 pressures between 0 and \SI{31}{GPa} with $K = 200$ live walkers per pressure.
Configurations are exchanged between neighbouring pressures after every NS iteration, and the runs terminate when the NS temperature \cite{baldock_constant-pressure_2017} reaches \SI{100}{K}. The move set, which includes Zn--S swap moves, and all further settings are listed in the SI.
The run takes \num{43608} NS iterations and $1.5\times10^{9}$ energy evaluations of the foundation model.

The heat capacity and density of this run (Fig.~\ref{fig:re_acc}a,b) already show the overall picture.
A melting ridge in $C_p$ lies near \SI{2000}{K} at low pressure (\SI{1980}{K} at \SI{0}{GPa}), and a step in density marks the B3$\to$B1 transition at \SIrange{17}{18}{GPa} at low temperature, moving to \SIrange{15}{16}{GPa} at \SI{1500}{K}.
The low-temperature density at \SI{0}{GPa}, \SI{4.10}{g\,cm^{-3}}, agrees with the measured \SI{4.13}{g\,cm^{-3}}~\cite{vasilyeva1998}.
In the 16-atom cell, however, the melting peak is broad ($C_p/(N k_\mathrm{B}) \leq 9$), and the solid--solid boundary is resolved only on the \SI{1}{GPa} pressure grid.

The exchange between neighbouring pressures stays active over almost the whole run. 
Fig.~\ref{fig:re_acc}c shows the total RE acceptance rates resolved for each neighbouring replica pair and iteration, arising from a total of $n_\mathrm{RE}=50$ swap attempts, see Eq.~\eqref{eq:n_re}.
Per pair, 16--63\,\% of the attempted swaps are accepted on average, and the median over the pairs stays above 0.5 for the first half of the run.
It decreases only towards the end, as the replicas settle into structurally distinct low-enthalpy configurations.
The acceptance falls below 1\,\% only for the pair at \SIrange{17}{18}{GPa}, after crossing the B3$\to$B1 transition, from iteration \num{33000} on, and for three further pairs within the last \num{1200} iterations.
This loss of overlap only biases a replica's predictions if it compromises sampling of the correct constrained prior distribution. 
In the present example, no bias arises if both the 17 and \SI{18}{GPa} replicas already correctly sample the B3–B1 mixture.

\begin{figure}[htbp]
  \includegraphics*[width=0.8\columnwidth]{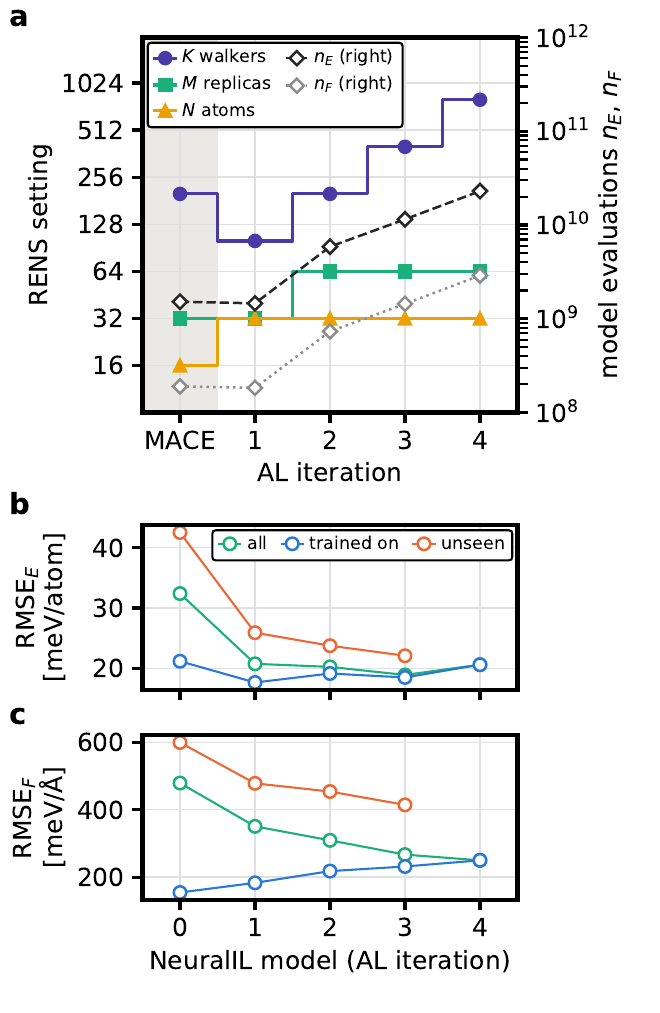}
  \caption{Active learning of the \software{NeuralIL} committee.
    (a)~Settings of the nested-sampling run of each iteration: walkers per replica $K$, replicas $M$ and atoms $N$ (left axis), and the number of energy ($n_E$) and force ($n_F$) evaluations of the model (right axis).
    Iteration~0 is the foundation-model run of Fig.~\ref{fig:re_acc}; the run of iteration $k \geq 1$ uses the committee trained after iteration $k-1$.
    (b,c)~Energy and force RMSE against r\textsuperscript{2}SCAN of the committee trained in each iteration, on all DFT sets of the run, on the sets it was trained on, and on the later sets it has not seen.}
  \label{fig:al}
\end{figure}

\begin{figure*}
  \centering
  \includegraphics[width=\textwidth]{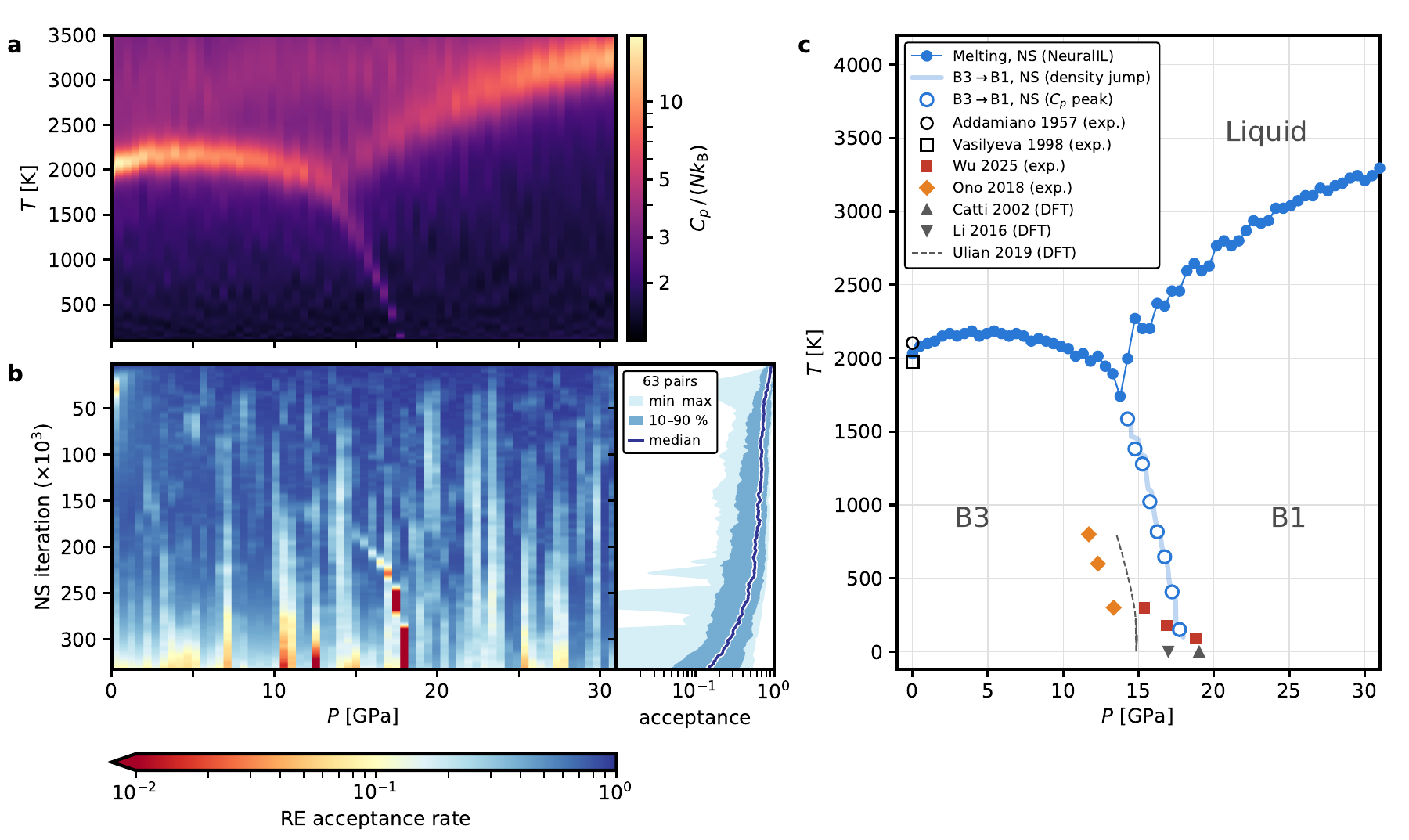}
  \caption{
    \textbf{RENS prediction of the ZnS phase diagram.}
    (a)~Heat capacity $C_p/(N k_\mathrm{B})$.
    (b)~Replica-exchange acceptance rate as in Fig.~\ref{fig:re_acc}c, for the 63 pairs of neighbouring pressures.
    (c)~Phase diagram: melting line from the $C_p$ peaks (filled circles), B3$\to$B1 boundary from the density jump (shaded line) and its $C_p$ peaks (open circles), compared with the ambient-pressure melting points~\cite{addamiano1957,vasilyeva1998} and with experimental~\cite{wu2025,ono2018} and DFT~\cite{catti2002,li2016,ulian2019} B3$\to$B1 transition pressures.}
  \label{fig:zns}
\end{figure*}

\subsubsection{Distillation and active learning refinement}

Resolving the phase boundaries requires larger cells and denser pressure grids than the foundation model affords, so we distil the foundation-model into a compact, system-specific potential and refine it by active learning.
From the 32 replica trajectories of the foundation-model run we stride-sample 1000 configurations, label them with r\textsuperscript{2}SCAN density functional theory (see SI for details), and train a committee of ten \software{NeuralIL} models~\cite{neuralil} with a \SI{5}{\angstrom} cutoff on a 90/10 train/validation split.
The foundation model thus only selects the configurations, while the labels come from the functional it was trained on.
The cheaper \software{NeuralIL} model allows NS for a larger cell of 32 atoms, which we refine by the active learning strategy proposed in Ref.~\cite{unglert_active_2026}.
At each iteration, NS is run with the current committee, and \num{10000} configurations are sampled at regular intervals from the resulting trajectories, leveraging the hierarchical sampling inherent to NS to promote diversity.
The 200 configurations with the highest committee force uncertainty,\cite{Heid_JCIM24} measured relative to the force magnitude, are then labelled with DFT, and the committee is retrained from scratch on all labels accumulated so far.
Over four iterations the NS settings are ramped from 32 pressures with $K = 100$ to 64 pressures with $K = 200$, 400 and 800 (Fig.~\ref{fig:al}a), raising the cost of a run from $1.5\times10^{9}$ to $2.3\times10^{10}$ energy evaluations, and 1792 configurations are labelled in total.

Figure~\ref{fig:al}b,c follows the accuracy of the \software{NeuralIL} committees over the active-learning iterations.
We score the committee of every iteration on every DFT set collected in the run and split the sets into those it was trained on -- the 1000 configurations from the foundation-model run and the sets of all iterations up to its own -- and the later ones it has never seen.
On the unseen sets the error drops with every iteration, from \SI{42}{} to \SI{22}{meV/atom} in energy and from \SI{0.60}{} to \SI{0.41}{eV/\angstrom} in force.
On the training sets it stays at \SIrange{18}{21}{meV/atom} and \SIrange{0.15}{0.25}{eV/\angstrom}, the force error rising slightly as the configurations of the \software{NeuralIL} runs, sampled at harsher conditions in the larger cell, enter the training set.
The last committee reaches \SI{21}{meV/atom} and \SI{0.25}{eV/\angstrom} on all 1792 configurations.

\subsubsection{The final phase diagram}

Figure~\ref{fig:zns} shows the final NS run, with the committee of the third iteration (trained on 1600 configurations), at 64 pressures between 0 and \SI{31}{GPa}. 
With $K = 800$ it takes \num{334443} NS iterations (Table~S8), resulting in the total of $2.3\times10^{10}$ model evaluations.
To put this computational effort into perspective, we measured the average model evaluation time using a batch of 512 configurations extracted from the NS trajectories.
Extrapolating these timings to the full simulation yields an estimated runtime of 2.2 years for a serial ASE-based NS driver (compare the orange and red points in Fig.~\ref{fig:scaling}).
In contrast, the simulation with \software{JAXRENS} completed in just 5.8 days on four GPUs.
Compared with the foundation-model run, the melting peak in the larger cell is considerably sharper (Fig.~\ref{fig:zns}a), as expected from the reduced finite-size broadening of a first-order transition.
The low-temperature density at \SI{0}{GPa}, \SI{4.14}{g\,cm^{-3}}, again agrees with experiment~\cite{vasilyeva1998}.
Due to the larger $K$, the statistical noise in $C_p$ below the melting transition is considerably reduced compared to the foundation model simulation. 
Despite the larger cell, the replica exchange is even more effective than in the foundation-model run (Fig.~\ref{fig:zns}b), with 30--77\,\% of the swaps accepted per pair, consistent with the closer spacing of the pressure grid.
The B3$\to$B1 transition shows up as a diagonal of reduced acceptance that follows the boundary to higher pressure as the temperature decreases.
Only the two pairs between \SI{17.2}{} and \SI{18.2}{GPa}, from iteration \num{248000} on, and one pair at \SIrange{12.3}{12.8}{GPa} near the end of the run fall below 1\,\% acceptance.

Figure~\ref{fig:zns}c collects the transitions extracted from the final \software{NeuralIL} run.
For each pressure, the melting temperature is taken as the lowest-temperature peak of $C_p(T)$ above \SI{1500}{K} with a prominence of at least $0.2\,k_\mathrm{B}$ per atom.
The tetrahedral B3 phase melts at \SIrange{2031}{2184}{K} up to \SI{10}{GPa}.
Its melting temperature then drops steeply to \SI{1740}{K} at \SI{13.8}{GPa}, close to the B3--B1--liquid triple point, beyond which the rocksalt melting line rises from about \SI{2200}{K} at \SIrange{14.8}{15.8}{GPa} to \SI{3295}{K} at \SI{31}{GPa}.
Just above the triple point, at \SI{14.8}{GPa}, the melting peak is weak, but it lies in line with the neighbouring pressures and in the temperature range where the density drops.
A more detailed assignment of the transition points can be found in the SI.
At ambient pressure the computed \SI{2031}{K} lies between the melting point of stoichiometric ZnS (\SI{1973}{K})~\cite{vasilyeva1998} and the frequently cited \SI{2103}{K} of Ref.~\cite{addamiano1957}, which Ref.~\cite{vasilyeva1998} assigns to a Zn-rich composition.

The B3$\to$B1 transition leaves only a weaker signature in $C_p$, a consequence of the smaller latent heat compared to the liquid$\to$solid transition.
Thanks to the replica exchange it is nevertheless perfectly resolved at a pressure spacing of \SI{0.5}{GPa}: eight neighbouring replicas between \SI{14.3}{} and \SI{17.7}{GPa} show a $C_p$ peak at which the density jumps, reflecting the tetrahedral-to-octahedral transition.
The boundary runs from \SI{18.0}{GPa} at \SI{100}{K} to \SI{14.5}{GPa} at \SI{1590}{K}, the highest temperature at which we evaluate it, close to the triple point.
The negative Clapeyron slope of $P(\mathrm{GPa}) \approx 18.2 - 0.0023\,T(\mathrm{K})$, obtained by a linear fit to the density jump, is in good agreement with experiment~\cite{ono2018} and with the quasi-harmonic DFT calculations of Ulian \textit{et al.}~\cite{ulian2019}.
The absolute transition pressure lies \SIrange{4}{5}{GPa} above the high-temperature X-ray data of Ono \textit{et al.}~\cite{ono2018}.
This offset is inherited from the \SI{0}{K} r\textsuperscript{2}SCAN transition pressure of \SI{17.9}{GPa}, which lies within the range of previous DFT results (\SIrange{14.7}{19}{GPa}~\cite{catti2002,li2016,ulian2019}) and close to the \SI{0}{K} extrapolation of the low-temperature Raman data of Wu \textit{et al.} (\SI{20.3}{GPa}~\cite{wu2025}), but above the extrapolation of the high-temperature diffraction data (\SI{14.4}{GPa}~\cite{ono2018}).

\section{Conclusion}
\label{sec:conclusion}

We have presented \software{JAXRENS}, a GPU-native, massively parallel NS framework that removes the practical bottlenecks separating atomistic NS from modern MLIPs. 
By expressing the entire sampling computation as composable \software{JAX} transformations, the framework advances a vectorised ensemble of walkers in a single JIT-compiled step, shards a single large run across multiple GPUs, and distributes ensembles of replica-exchange-coupled runs across devices, all without bespoke kernels.
Benchmarks against CPU-oriented codes show that this design substantially raises the throughput of atomistic NS on accelerator hardware (Sec.~\ref{sec:scaling}), and a pluggable backend interface lets us drive the same sampler with the \software{MACE}, \software{NeuralIL}, and \software{Nequix} potentials without modifying the sampling code.

We have also demonstrated what this enables on a practical materials-science problem: the pressure--temperature phase diagram of ZnS up to \SI{31}{GPa}, including a melting line that has not been measured under pressure.
Starting from the \software{MACE-MATPES-r2SCAN-0} foundation model, NS of a 16-atom cell provided both a first phase diagram and the configurations from which a compact \software{NeuralIL} potential was trained on r$^2$SCAN single points; the cheaper surrogate in turn made sampling of a 32-atom cell affordable, and a few active-learning iterations on that cell extended its training set to adjust the model for the harsh finite-temperature configurations faced in NS. 
The sampled ambient-pressure melting temperature of 2031~K lies between the measured melting points of stoichiometric and of Zn-rich ZnS, and the computed zincblende--rocksalt boundary shows the experimentally observed negative Clapeyron slope, its absolute position being inherited from the r$^2$SCAN transition pressure rather than from the sampling. 
Beyond the available experiments, the calculations predict a minimum of the melting temperature near the zincblende--rocksalt--liquid triple point at about 14~GPa and a steeply rising rocksalt melting line above it.
No phases were prescribed to the sampler: the solid--solid and melting transitions emerge directly from the heat capacity and density of the sampled partition function. 
To our knowledge this is the first NS study driven by a machine-learned foundation model, and it reduces the system-specific input of a first-principles-quality phase diagram to a limited number of reference calculations.
Together with concurrent efforts to probe the finite-temperature capabilities of foundation models through automated active learning~\cite{gawkowski_dyna-mat_2026}, this points towards finite-temperature phase stability becoming a routine property of a foundation model.

Several directions follow naturally.
The ZnS workflow already shows the benefit of handing the expensive sampling from the foundation model to a compact, system-specific surrogate, which our benchmarks show exposes far more of the framework's batching headroom.
Labelling the surrogate's training data with the foundation model itself instead of DFT would remove the remaining DFT step and could yield large additional speed-ups at retained accuracy.
The resulting savings enable larger simulation cells and denser pressure grids, helping to reduce finite-size broadening and sharpen the extracted phase boundaries while also bringing more complex systems within reach.
Finally, the distributed replica-exchange machinery is not specific to binary compounds: the same infrastructure extends to multicomponent systems, where the combinatorial growth of accessible phases makes automatic, order-parameter-free phase discovery most valuable.
We view this work as an infrastructural step towards routine, first-principles-quality finite-temperature phase stability prediction with machine-learned atomistic models.

\section{Methods}
\label{sec:methods}

\subsection{Nested sampling}
\label{sec:meth_ns}

In the following, we develop the formalism implemented in \software{JAXRENS} for the thermodynamic $NPT$ ensemble. Analogous formulations apply to related ensembles, such as the sGC-$\mu$PT ensemble. The targeted isothermal--isobaric partition function \cite{baldock2016} reads
\begin{equation}
\begin{aligned}
    Z(\beta;P) \propto
    &\int_0^\infty \!\mathrm{d}V
    \int \!\mathrm{d}\mathbf{h}_0\,
    \delta(\det\mathbf{h}_0 - 1) \\
    &\times \int_{D(V)} \!\mathrm{d}\mathbf{q}\;
    \exp\!\left\{
        -\beta\left[
            U(\mathbf{q};V,\mathbf{h}_0) + PV
        \right]
    \right\}.
\end{aligned}
\label{eq:zdef}
\end{equation}
where $\beta = 1/k_\mathrm{B}T$, $V$ denotes the cell volume, and $\mathbf{h}_0$ is the unit-determinant cell matrix describing its shape, such that $\mathbf{h}=V^{1/3}\mathbf{h}_0$.
The Cartesian atomic coordinates are collected in $\mathbf{q}$, with $D(V)$ denoting the corresponding configuration domain. 
The momentum degrees of freedom have been integrated out analytically.
In practice, we express atomic positions in scaled coordinates, $\mathbf{q}_j=\mathbf{h}\mathbf{s}_j$, with $\mathbf{s}_j\in[0,1)^3$.
We therefore write $\mathbf{r}=(V,\mathbf{h}_0,\mathbf{s})$ for the complete configuration, comprising the cell volume, shape, and scaled atomic coordinates.

The difficulty is that this configurational integral is dominated by exponentially small regions of configuration space, which conventional temperature-based sampling only locates if the relevant phases are known and targeted in advance.
Nested sampling \cite{skilling2006,partay2010,baldock2016} evaluates the integral instead by scanning the configuration space from high to low enthalpy, without reference to any particular temperature.
It maintains a live set of $K$ walkers distributed according to the prior $\pi(\mathbf{r})$ restricted to configurations below a running enthalpy threshold $H_i$,
\begin{equation}
    f_i(\mathbf{r})
    =
    \frac{
        \pi(\mathbf{r})
        \,\mathbb{I}\!\left[
            U(\mathbf{r}) + P V(\mathbf{r})
            <
            H_i
        \right]
    }{
        X_i
    },
    \label{eq:constrained-prior}
\end{equation}
where $\mathbb{I}$ is the indicator function and the normalisation $X_i = \int \pi(\mathbf{r})\,\mathbb{I}[U + PV < H_i]\,\mathrm{d}\mathbf{r}$ is the \emph{prior mass} enclosed by the threshold.
According to the principle of \textit{equal a priori probability}, $\pi(\mathbf{r})$ in general needs to be uniform. 
At each iteration the walker with the highest enthalpy is removed and its enthalpy becomes the new threshold $H_i$; the removed walker is replaced by a clone of a surviving one, which is then decorrelated under the tightened constraint.
Because the discarded point is, by construction, the highest of $K$ samples drawn uniformly in prior mass, the enclosed volume contracts by the known statistical factor $\frac{K}{K+1}$ at every iteration, $X_i = \big(\frac{K}{K+1}\big)^i \simeq \exp(-i/K)$, so the multi-dimensional integral collapses onto the sequence of recorded enthalpies.
The partition function at any temperature then follows from a one-dimensional quadrature,
\begin{equation}
    Z(\beta;P)
    =
    \sum_i
    \left(
        X_{i-1}
        -
        X_i
    \right)
    \exp\!\left[
        -\beta H(\mathbf{r}_i)
    \right],
    \label{eq:partition}
\end{equation}
with $H(\mathbf{r}_i) = U(\mathbf{r}_i) + P V(\mathbf{r}_i)$, from which all equilibrium thermodynamic observables are obtained by standard differentiation.
A single run therefore yields the full temperature dependence, and phase transitions appear as peaks in the resulting heat capacity without any order parameter having been specified beforehand.

The one step that is not analytic is the decorrelation of the cloned walker, which must produce an effectively independent sample from the constrained prior of Eq.~\eqref{eq:constrained-prior}.
We generate it with an MCMC transition kernel $\mathcal{K}_i$ that leaves $f_i$ invariant and is composed in a Metropolis-within-Gibbs fashion following Ref.~\cite{baldock2016}. 
It is built from a set of $n_\mathrm{steptypes}$ elementary move kernels $\{\kappa_k\}_{k=1}^{n_\mathrm{steptypes}}$ with weights $\{w_k\}$: at every MCMC step a single kernel is drawn with probability $p_k = w_k / \sum_j w_j$ and applied to the walker.
The available kernels act on the atomic positions -- random-walk, Galilean, and Hamiltonian Monte Carlo updates -- and on the simulation cell -- volume, stretch, and shear moves -- so that the $NPT$ ensemble is sampled by including the cell moves in the mixture.
Since the constraint enters as a hard threshold rather than a Boltzmann weight, a proposal that is symmetric with respect to the prior is simply accepted whenever it stays below the current threshold,
\begin{equation}
    \alpha(\mathbf{r} \to \mathbf{r}')
    =
    \mathbb{I}\!\left[
        U(\mathbf{r}') + P V(\mathbf{r}')
        <
        H_i
    \right],
\end{equation}
with the volume move carrying an additional Jacobian factor to compensate its non-symmetric proposal.
The step size of each kernel is adapted during the run so that its acceptance rate stays within a prescribed window ($0.2$--$0.5$ in the ZnS runs of Sec.~\ref{sec:zns}).
Since this decorrelation has to be repeated for every discarded walker throughout the simulation, and each MCMC step requires an energy evaluation, it constitutes the dominant computational cost of nested sampling and the principal target of the parallelisation described in Sec.~\ref{sec:overview}.

\subsection{Replica-exchange nested sampling}
\label{sec:meth_rens}

We use the RENS method proposed in Ref.~\cite{unglert_replica_2025}. Therein, replica-exchange (RE) moves couple a set of $M$ NS simulations at different external pressures $\{P_m\}_{m=1}^M$, simultaneously sampling energy-constrained prior distributions $f_i^{[m]}(\mathbf{r})$ at iteration $i$. In this fashion, a RENS simulation samples the configurational density of states at each pressure with strongly enhanced ergodicity.

For replica $m$, the constrained prior at NS iteration $i$ is given by
\begin{equation}
    f_i^{[m]}(\mathbf{r})
    =
    \frac{
        \pi(\mathbf{r})
        \,\mathbb{I}\!\left[
            U(\mathbf{r}) + P_m V(\mathbf{r})
            <
            H_i^{[m]}
        \right]
    }{
        X_i^{[m]}
    },
\end{equation}
where $\pi(\mathbf{r})$ denotes the prior distribution, $H_i^{[m]}$ is the current enthalpy constraint, $X_i^{[m]}$ is the corresponding prior mass, and $\mathbb{I}$ is the indicator function. 
After removing the live point with the highest enthalpy, a surviving live point is cloned and evolved using an MCMC transition kernel $\mathcal{K}_i^{[m]}$ that leaves the constrained prior invariant.
The complex energy landscapes commonly encountered in materials applications make constrained target distributions difficult to sample, often leading to broken ergodicity in independent NS simulations.
RE moves alleviate this problem by allowing configurations to diffuse between pressures, thereby substantially improving sampling efficiency.

Introducing the product state
\begin{equation}
    \mathbf{R}
    =
    \left(
        \mathbf{r}^{[1]},
        \ldots,
        \mathbf{r}^{[M]}
    \right),
\end{equation}
the local constrained MCMC updates define the product kernel
\begin{equation}
    \mathcal{K}_i^{\mathrm{loc}}
    =
    \bigotimes_{m=1}^{M}
    \mathcal{K}_i^{[m]},
\end{equation}
which independently propagates each replica while leaving the joint constrained prior
\begin{equation}
    f_i(\mathbf{R})
    =
    \prod_{m=1}^{M}
    f_i^{[m]}(\mathbf{r}^{[m]})
\end{equation}
invariant. Between successive NS iterations, the replicas are coupled by
RE kernels acting on the product state. We define the
elementary swap kernel $\mathcal{S}_{m,m+1}$, which proposes exchanging
the neighbouring replicas $(m,m+1)$ and accepts the move with probability
\begin{equation}
    \alpha_{m,m+1}
    =
    \min\left\{
        1,
        \frac{
            f_i^{[m]}(\mathbf{r}^{[m+1]})
            f_i^{[m+1]}(\mathbf{r}^{[m]})
        }{
            f_i^{[m]}(\mathbf{r}^{[m]})
            f_i^{[m+1]}(\mathbf{r}^{[m+1]})
        }
    \right\}.
\end{equation}
Independent swaps are executed simultaneously using the composite even
and odd kernels
\begin{equation}
    \mathcal{S}_{\mathrm{even}}
    =
    \prod_{m\ \mathrm{even}}
    \mathcal{S}_{m,m+1},
    \qquad
    \mathcal{S}_{\mathrm{odd}}
    =
    \prod_{m\ \mathrm{odd}}
    \mathcal{S}_{m,m+1},
\end{equation}
which together define a replica-exchange cycle
\begin{equation}
    \label{eq:n_re}
    \mathcal{S}
    =
    \left(
        \mathcal{S}_{\mathrm{odd}}
        \circ
        \mathcal{S}_{\mathrm{even}}
    \right)^{n_{\mathrm{RE}}},
\end{equation}
where $n_{\mathrm{RE}}$ denotes the number of even-odd swap rounds
performed between two successive NS iterations.

The partition functions $\{Z(\beta;P_m)\}_{m=1}^M$ are then obtained by applying the thermal reweighting of Eq.~\eqref{eq:partition} to the density of states sampled by each replica, using its own prior masses $X^{[m]}_i$ and enthalpies $H(\mathbf{r}^{[m]}_i) = U(\mathbf{r}^{[m]}_i) + P_m V(\mathbf{r}^{[m]}_i)$.
A single RENS run thus delivers the thermodynamics over the whole temperature range at every pressure in the set.

\subsection{Timing benchmarks}
\label{sec:meth_timing}

We benchmark \software{JAXRENS}, using either the \software{JAX} implementation of \software{MACE} (\software{mace-jax})~\cite{batatia2022mace,batatia_design_2025} or \software{NeuralIL}~\cite{neuralil}, against \software{pymatnest}~\cite{pymatnest} and \software{FreeBird.jl}~\cite{freebird}, both of which access the PyTorch implementation (\software{mace-torch}) through its ASE calculator. 
All three drivers use the same \software{MACE-MP-0} ``small'' weights~\cite{batatia2025foundation} and are initialised from identical rattled diamond configurations of Si$_N$ ($N = 16$, 32, 64) at volumes of 54 or 27\,\AA$^3$/atom. 
Figure~\ref{fig:scaling} shows the results for Si$_{32}$ at 27\,\AA$^3$/atom. 
Walker decorrelation uses atom-only Galilean Monte Carlo (GMC) in \software{JAXRENS} and \software{pymatnest}, and single-atom random walks in \software{FreeBird.jl}. 
GPU benchmarks run on NVIDIA H100 GPUs, while CPU benchmarks use one AMD EPYC~9654 core per MPI rank. 
Full driver settings, results for all six systems, and software versions are provided in the SI.

The cost unit is one \emph{model call}, a single evaluation of one configuration.
Because the drivers bundle different numbers of calls into one step or iteration, we count the calls executed in every run and report the measured step or iteration time divided by that count.
The comparison is set up to favour the existing codes: their calculators compute only what the sampler requests (mostly the energy alone), whereas \software{JAXRENS} evaluates both energies and forces in every GMC call and pads its neighbour lists to a fixed size. These additional costs arise from the static-shape, compiled-execution paradigm of \software{JAX}, which enables efficient execution on hardware accelerators.
We time the elementary NS step in isolation, after discarding warm-up steps, and take the median over the timed steps of a run; every measurement is repeated five times as independent processes with different random seeds, and we report the mean and standard deviation over the repeats.

Parallel efficiency is measured at a fixed budget of $\approx~\!1000$ model calls per step, distributed over an increasing parallel width: the number $n_\mathrm{walked}\in\{1,2,\dots,256\}$ of walkers evaluated as one batch in \software{JAXRENS}, and the number of MPI ranks ($1$--$16$) for \software{pymatnest}, whose budget is split across ranks.
Beyond one GPU, the walkers of a single run are sharded over up to four GPUs.
To attribute the cost of a single call, one energy-and-force evaluation is additionally timed and profiled outside any sampler in both \software{MACE} implementations.

Details on the training of the distilled \software{MACE} and the \software{NeuralIL} surrogate model can be found in the SI.

\subsection{AI and LLM usage}
\label{sec:meth_llm}

OpenAI GPT-6 was used to assist with language editing of the manuscript.
Claude Opus 5 was used to assist with source code development, testing, and documentation of \software{JAXRENS}, as well as the preparation of performance benchmarks.
The authors take full responsibility for the final content.

\section*{Code availability}

\texttt{JAXRENS} is openly available at
\url{https://github.com/nunglert/jaxrens}, together with extensive documentation,
installation instructions, and worked examples at \url{https://nunglert.github.io/jaxrens/}. 

\section*{Data availability}

The benchmark scripts, container definitions and timing data as well as the input files for the ZnS simulations are available in a separate zenodo repository: \url{https://doi.org/10.5281/zenodo.24276386}.
To ensure reproducibility, each driver used was executed within an Apptainer container, with the corresponding images also provided.
In addition we provide Supporting Information with more detailed information on the conducted numerical experiments.

\section*{Author contributions}
N.U. developed the code, designed the workflow, generated and analysed all data, and wrote the first draft of the manuscript. 
A.S. contributed to the development and testing of JAXRENS. 
J.C. contributed to the early development of the code, provided feedback throughout, and contributed to the writing. 
G.K.H.M. supervised the project, secured funding, and contributed to the writing. 
All authors reviewed the manuscript.

\section*{Acknowledgments}
This research was funded in part by the Austrian Science Fund (FWF) through 10.55776/F81 and 10.55776/COE5. 
J.C. furthermore acknowledges Grant CEX2023-001286-S funded by MICIU/AEI /10.13039/501100011033, and funding from  the Spanish MICIU/AEI /10.13039/501100011033 and the European union FEDER through project PID2023-148359NB-C21.
Parts of the computational results have been achieved using the Austrian Scientific Computing (ASC) infrastructure.
For open access purposes, the authors have applied a CC BY public copyright licence to any author accepted manuscript version arising from this submission.

\bibliography{refs}

@article{Heid_JCIM24,
    author = {Heid, Esther and Schörghuber, Johannes and Wanzenböck, Ralf and Madsen, Georg K. H.},
    title = {Spatially Resolved Uncertainties for Machine Learning Potentials},
    journal = {J. Chem. Inf. Model.},
    volume = {64},
    pages = {6377-6387},
    year = {2024},
    doi = {10.1021/acs.jcim.4c00904},
}

@software{JAX,
  author = {James Bradbury and Roy Frostig and Peter Hawkins and Matthew James Johnson and Yash Katariya and Chris Leary and Dougal Maclaurin and George Necula and Adam Paszke and Jake Vander{P}las and Skye Wanderman-{M}ilne and Qiao Zhang},
  title = {{JAX}: composable transformations of {P}ython+{N}um{P}y programs},
  url = {http://github.com/jax-ml/jax},
  version = {0.3.13},
  year = {2018},
}

@misc{Riccius2026,
      title={Finite-Temperature Thermodynamics of Cu(100) Oxidation: Missing-Row Reconstruction, Defect States, and Order-Disorder Transition from Nested Sampling}, 
      author={Felix Riccius and Karsten Reuter and Hendrik H. Heenen and Jutta Rogal},
      year={2026},
      eprint={2608.12787},
      archivePrefix={arXiv},
      primaryClass={cond-mat.mtrl-sci},
      url={https://arxiv.org/abs/2608.12787}, 
}

@article{skilling2006,
  author  = {Skilling, John},
  title   = {Nested Sampling for General {Bayesian} Computation},
  journal = {Bayesian Anal.},
  volume  = {1},
  number  = {4},
  pages   = {833--859},
  year    = {2006},
  doi     = {10.1214/06-BA127},
  url     = {https://doi.org/10.1214/06-BA127},
}

@article{partay2010,
  author  = {P{\'a}rtay, Livia B. and Bart{\'o}k, Albert P. and Cs{\'a}nyi, G{\'a}bor},
  title   = {Efficient Sampling of Atomic Configurational Spaces},
  journal = {J. Phys. Chem. B},
  volume  = {114},
  number  = {32},
  pages   = {10502--10512},
  year    = {2010},
  doi     = {10.1021/jp1012973},
  url     = {https://doi.org/10.1021/jp1012973},
}

@article{baldock2016,
  author  = {Baldock, Robert J. N. and P{\'a}rtay, Livia B. and
             Bart{\'o}k, Albert P. and Payne, Mike C. and Cs{\'a}nyi, G{\'a}bor},
  title   = {Determining Pressure-Temperature Phase Diagrams of Materials},
  journal = {Phys. Rev. B},
  volume  = {93},
  number  = {17},
  pages   = {174108},
  year    = {2016},
  doi     = {10.1103/PhysRevB.93.174108},
  url     = {https://doi.org/10.1103/PhysRevB.93.174108},
}

@misc{pymatnest,
  author       = {Bernstein, Noam and Cs{\'a}nyi, G{\'a}bor and P{\'a}rtay, Livia B.},
  title        = {{pymatnest}: A Software Library for Nested Sampling of Materials},
  year         = {2016},
  howpublished = {\url{https://github.com/libAtoms/pymatnest}},
  url          = {https://github.com/libAtoms/pymatnest},
}

@article{freebird,
  author  = {Yang, Ray and Chen, Junchi and Thibodeaux, Douglas and
             Wexler, Robert B.},
  title   = {{FreeBird.jl}: An Extensible Toolbox for Simulating Interfacial
             Phase Equilibria},
  journal = {J. Chem. Theory Comput.},
  volume  = {21},
  number  = {21},
  pages   = {10765--10779},
  year    = {2025},
  doi     = {10.1021/acs.jctc.5c01348},
  url     = {https://doi.org/10.1021/acs.jctc.5c01348},
}

@article{unglert_replica_2025,
  author  = {Unglert, N. and P{\'a}rtay, L. B. and Madsen, G. K. H.},
  title   = {Replica Exchange Nested Sampling},
  journal = {J. Chem. Theory Comput.},
  volume  = {21},
  number  = {15},
  pages   = {7304--7319},
  year    = {2025},
  doi     = {10.1021/acs.jctc.5c00588},
  url     = {https://doi.org/10.1021/acs.jctc.5c00588},
}

@article{unglert_active_2026,
  author  = {Unglert, Nico and Ketter, Michael and Madsen, Georg K. H.},
  title   = {Active Learning Potentials for First-Principles Phase Diagrams
             Using Replica-Exchange Nested Sampling},
  journal = {npj Comput. Mater.},
  volume  = {12},
  number  = {1},
  pages   = {107},
  year    = {2026},
  doi     = {10.1038/s41524-026-01989-z},
  url     = {https://doi.org/10.1038/s41524-026-01989-z},
}

@article{fletcher2025,
  author  = {Fletcher, Vincent G. and Bart{\'o}k, Albert P. and P{\'a}rtay, Livia B.},
  title   = {Autonomous Thermodynamically Informed Database Generation for
             Machine-Learned Interatomic Potentials and Application to Magnesium},
  journal = {npj Comput. Mater.},
  volume  = {12},
  number  = {1},
  pages   = {36},
  year    = {2025},
  doi     = {10.1038/s41524-025-01903-z},
  url     = {https://doi.org/10.1038/s41524-025-01903-z},
}

@misc{hellyar2026,
  author        = {Hellyar, Tom and Salzbrenner, Pascal T. and
                   Cooke, Peter I. C. and Pickard, Chris J. and
                   Habershon, Scott and P{\'a}rtay, Livia B.},
  title         = {Benchmarking Empirical and Machine-Learned Interatomic
                   Potentials Using Phase Diagram Predictions for Lead},
  year          = {2026},
  eprint        = {2605.16018},
  archivePrefix = {arXiv},
  primaryClass  = {cond-mat.mtrl-sci},
  url           = {https://arxiv.org/abs/2605.16018},
}

@article{marchant2023carbon,
  author    = {Marchant, George A. and Caro, Miguel A. and
               Karasulu, Bora and P{\'a}rtay, Livia B.},
  title     = {Exploring the Configuration Space of Elemental Carbon with
               Empirical and Machine Learned Interatomic Potentials},
  journal   = {npj Comput. Mater.},
  volume    = {9},
  number    = {1},
  pages     = {131},
  year      = {2023},
  doi       = {10.1038/s41524-023-01081-w},
  url       = {https://doi.org/10.1038/s41524-023-01081-w},
}

@article{unglert2023silicon,
  author  = {Unglert, Nico and Carrete, Jes{\'u}s and P{\'a}rtay, Livia B. and
             Madsen, Georg K. H.},
  title   = {Neural-Network Force Field Backed Nested Sampling:
             Study of the Silicon $p$--$T$ Phase Diagram},
  journal = {Phys. Rev. Mater.},
  volume  = {7},
  number  = {12},
  pages   = {123804},
  year    = {2023},
  doi     = {10.1103/PhysRevMaterials.7.123804},
  url     = {https://doi.org/10.1103/PhysRevMaterials.7.123804},
}

@inproceedings{batatia2022mace,
  author        = {Batatia, Ilyes and Kov{\'a}cs, D{\'a}vid P{\'e}ter and
                   Simm, Gregor N. C. and Ortner, Christoph and Cs{\'a}nyi, G{\'a}bor},
  title         = {{MACE}: Higher Order Equivariant Message Passing Neural Networks
                   for Fast and Accurate Force Fields},
  booktitle     = {Advances in Neural Information Processing Systems},
  volume        = {35},
  pages         = {11423--11436},
  year          = {2022},
  publisher     = {Curran Associates, Inc.},
  eprint        = {2206.07697},
  archivePrefix = {arXiv},
  primaryClass  = {stat.ML},
  url           = {https://arxiv.org/abs/2206.07697},
}

@article{batzner2022nequip,
  author  = {Batzner, Simon and Musaelian, Albert and Sun, Lixin and
             Geiger, Mario and Mailoa, Jonathan P. and Kornbluth, Mordechai and
             Molinari, Nicola and Smidt, Tess E. and Kozinsky, Boris},
  title   = {{E(3)}-Equivariant Graph Neural Networks for Data-Efficient
             and Accurate Interatomic Potentials},
  journal = {Nat. Commun.},
  volume  = {13},
  number  = {1},
  pages   = {2453},
  year    = {2022},
  doi     = {10.1038/s41467-022-29939-5},
  url     = {https://doi.org/10.1038/s41467-022-29939-5},
}

@misc{nequix,
  author        = {Koker, Teddy and Kotak, Mit and Smidt, Tess},
  title         = {Training a Foundation Model for Materials on a Budget},
  year          = {2025},
  eprint        = {2508.16067},
  archivePrefix = {arXiv},
  primaryClass  = {cond-mat.mtrl-sci},
  doi           = {10.48550/arXiv.2508.16067},
  url           = {https://arxiv.org/abs/2508.16067},
  note          = {Code: \url{https://github.com/atomicarchitects/nequix}},
}

@article{batatia2025foundation,
  author  = {Batatia, Ilyes and Benner, Philipp and Chiang, Yuan and
             Elena, Alin M. and Kov{\'a}cs, D{\'a}vid P. and Riebesell, Janosh
             and others},
  title   = {A Foundation Model for Atomistic Materials Chemistry},
  journal = {J. Chem. Phys.},
  volume  = {163},
  number  = {18},
  pages   = {184110},
  year    = {2025},
  doi     = {10.1063/5.0297006},
  url     = {https://doi.org/10.1063/5.0297006},
}

@article{neuralil,
  author  = {Montes-Campos, Hadri{\'a}n and Carrete, Jes{\'u}s and
             Bichelmaier, Sebastian and Varela, Luis M. and Madsen, Georg K. H.},
  title   = {A Differentiable Neural-Network Force Field for Ionic Liquids},
  journal = {J. Chem. Inf. Model.},
  volume  = {62},
  number  = {1},
  pages   = {88--101},
  year    = {2022},
  doi     = {10.1021/acs.jcim.1c01380},
  url     = {https://doi.org/10.1021/acs.jcim.1c01380},
}

@inproceedings{jaxmd2020,
  author        = {Schoenholz, Samuel S. and Cubuk, Ekin D.},
  title         = {{JAX MD}: A Framework for Differentiable Physics},
  booktitle     = {Advances in Neural Information Processing Systems},
  volume        = {33},
  pages         = {11428--11441},
  year          = {2020},
  publisher     = {Curran Associates, Inc.},
  eprint        = {1912.04232},
  archivePrefix = {arXiv},
  primaryClass  = {physics.comp-ph},
  url           = {https://arxiv.org/abs/1912.04232},
}

@misc{torchsim2025,
  author        = {Gangan, Abhijeet S. and Riebesell, Janosh and
                   Goodall, Rhys E. A. and Cohen, Orion and others},
  title         = {{TorchSim}: An Efficient Atomistic Simulation Engine in {PyTorch}},
  year          = {2025},
  eprint        = {2508.06628},
  archivePrefix = {arXiv},
  primaryClass  = {physics.comp-ph},
  doi           = {10.48550/arXiv.2508.06628},
  url           = {https://arxiv.org/abs/2508.06628},
  note          = {Code: \url{https://github.com/Radical-AI/torch-sim}},
}

@misc{kups2026,
  author       = {Gao, Nicholas and K{\"o}hler, Jonas and Hanke, Felix and
                  Ramanan, Anita and Moubarak, Elias and Morrow, Joe and
                  de Haan, Pim and Openshaw, Hannah and Welling, Max and {CuspAI Team}},
  title        = {{kUPS}: A Universal Particle Simulation Toolkit},
  year         = {2026},
  publisher    = {CuspAI},
  howpublished = {\url{https://github.com/cusp-ai-oss/kups}},
  url          = {https://github.com/cusp-ai-oss/kups},
}

@misc{tojax2026,
  author       = {{Cusp AI}},
  title        = {{tojax}: Translating {PyTorch} Functions and Models to {JAX}},
  year         = {2026},
  publisher    = {CuspAI},
  howpublished = {\url{https://github.com/cusp-ai-oss/tojax}},
  url          = {https://github.com/cusp-ai-oss/tojax},
}

@misc{bradbury_jax_2018,
  title = {{JAX}: composable transformations of {Python}+{NumPy} programs},
  url = {http://github.com/google/jax},
  author = {Bradbury, James and Frostig, Roy and Hawkins, Peter and Johnson, Matthew James and Leary, Chris and Maclaurin, Dougal and Necula, George and Paszke, Adam and VanderPlas, Jake and Wanderman-Milne, Skye and Zhang, Qiao},
  year = {2018},
}

@misc{hu_demystifying_2026,
  title = {Demystifying {NCCL}: {An} {In}-depth {Analysis} of {GPU} {Communication} {Protocols} and {Algorithms}},
  shorttitle = {Demystifying {NCCL}},
  url = {http://arxiv.org/abs/2507.04786},
  doi = {10.48550/arXiv.2507.04786},
  urldate = {2026-05-29},
  publisher = {arXiv},
  author = {Hu, Zhiyi and Shen, Siyuan and Bonato, Tommaso and Jeaugey, Sylvain and Alexander, Cedell and Spada, Eric and Dinan, James and Hammond, Jeff and Hoefler, Torsten},
  month = mar,
  year = {2026},
  note = {arXiv:2507.04786 [cs.DC]},
}

@misc{gawkowski_dyna-mat_2026,
  title = {Dyna-{Mat}: {End}-to-end benchmarking of foundation machine learning interatomic potentials in finite-temperature ensembles},
  shorttitle = {Dyna-{Mat}},
  url = {http://arxiv.org/abs/2607.03433},
  doi = {10.48550/arXiv.2607.03433},
  urldate = {2026-07-13},
  publisher = {arXiv},
  author = {Gawkowski, Mikołaj J. and Artrith, Nongnuch and Bonfanti, Silvia and Gangan, Abhijeet Sadashiv and Heenen, Hendrik H. and Kioseoglou, Joseph and Lončarić, Ivor and Myneni, Hemanadhan and Riebesell, Janosh and Rossi, Mariana and Rupp, Matthias and Schmidt, Jonathan and Sharma, Shubham and Shi, Benjamin X. and Wadowski, Antoni and Hörmann, Lukas and Kapil, Venkat},
  month = jul,
  year = {2026},
  note = {arXiv:2607.03433 [cond-mat.mtrl-sci]
version: 1},
}

@article{furness_accurate_2020,
	title = {Accurate and {Numerically} {Efficient} r $^{\textrm{2}}$ {SCAN} {Meta}-{Generalized} {Gradient} {Approximation}},
	volume = {11},
	issn = {1948-7185, 1948-7185},
	url = {https://pubs.acs.org/doi/10.1021/acs.jpclett.0c02405},
	doi = {10.1021/acs.jpclett.0c02405},
	number = {19},
	urldate = {2023-05-02},
	journal = {J. Phys. Chem. Lett.},
	author = {Furness, James W. and Kaplan, Aaron D. and Ning, Jinliang and Perdew, John P. and Sun, Jianwei},
	month = oct,
	year = {2020},
	pages = {8208--8215},
}

@article{batatia_design_2025,
    title = {The design space of {E}(3)-equivariant atom-centred interatomic potentials},
    volume = {7},
    copyright = {2025 The Author(s)},
    issn = {2522-5839},
    url = {https://www.nature.com/articles/s42256-024-00956-x},
    doi = {10.1038/s42256-024-00956-x},
    number = {1},
    urldate = {2026-02-10},
    journal = {Nat. Mach. Intell.},
    publisher = {Nature Publishing Group},
    author = {Batatia, Ilyes and Batzner, Simon and Kovács, Dávid Péter and Musaelian, Albert and Simm, Gregor N. C. and Drautz, Ralf and Ortner, Christoph and Kozinsky, Boris and Csányi, Gábor},
    month = jan,
    year = {2025},
    pages = {56--67},
}

@article{addamiano1957,
  author  = {Addamiano, Arrigo and Dell, Paul A.},
  title   = {The Melting Point of Zinc Sulfide},
  journal = {J. Phys. Chem.},
  volume  = {61},
  number  = {7},
  pages   = {1020--1021},
  year    = {1957},
  doi = {10.1021/j150553a050}
}

@article{vasilyeva1998,
  author  = {Vasilyeva, I. G. and Belyaeva, E. I. and Gibner, Ya. I.},
  title   = {Determination of Melting Points for {A$^2$B$^6$} Compounds},
  journal = {J. Therm. Anal. Calorim.},
  volume  = {52},
  pages   = {403--412},
  year    = {1998},
  doi = {10.1023/A:1010118227837}
}

@article{ono2018,
  author  = {Ono, Shigeaki and Kikegawa, Takumi},
  title   = {Phase transition of {ZnS} at high pressures and temperatures},
  journal = {Phase Transit.},
  volume  = {91},
  number  = {1},
  pages   = {9--14},
  year    = {2018},
  doi     = {10.1080/01411594.2017.1350958},
}

@article{wu2025,
  author  = {Wu, Binbin and Lin, Yuru and Li, Yu and Chang, Xue and Tao, Yu
             and Liu, Jingyi and Lei, Li},
  title   = {Temperature effect of the pressure coefficients for multi-phonon
             coupling in {B3-ZnS}},
  journal = {Appl. Phys. Lett.},
  volume  = {126},
  pages   = {102103},
  year    = {2025},
  doi     = {10.1063/5.0258828},
}

@article{catti2002,
  author  = {Catti, Michele},
  title   = {First-principles study of the orthorhombic mechanism for the
             {B3/B1} high-pressure phase transition of {ZnS}},
  journal = {Phys. Rev. B},
  volume  = {65},
  pages   = {224115},
  year    = {2002},
  doi     = {10.1103/PhysRevB.65.224115},
}

@article{li2016,
  author  = {Li, Qiang and Zhang, Rui and Lv, Tianquan and Cao, Qilong},
  title   = {Theoretical study of {B3}-to-{B1} phase transition in {ZnS}},
  journal = {Phys. Lett. A},
  volume  = {380},
  pages   = {3696--3702},
  year    = {2016},
  doi     = {10.1016/j.physleta.2016.09.010},
}

@article{ulian2019,
  author  = {Ulian, Gianfranco and Valdr{\`e}, Giovanni},
  title   = {Thermomechanical, electronic and thermodynamic properties of
             {ZnS} cubic polymorphs: an ab initio investigation on the
             zinc-blende--rock-salt phase transition},
  journal = {Acta Crystallogr. Sect. B},
  volume  = {75},
  pages   = {1042--1059},
  year    = {2019},
  doi     = {10.1107/S2052520619012630},
}

@article{salomone_unbiased_2025,
  title = {Unbiased and consistent nested sampling via sequential {Monte} {Carlo}},
  issn = {1369-7412},
  url = {https://doi.org/10.1093/jrsssb/qkaf015},
  doi = {10.1093/jrsssb/qkaf015},
  urldate = {2025-08-21},
  journal = {J. R. Stat. Soc. Ser. B. Stat. Methodol.},
  author = {Salomone, Robert and South, Leah F and Drovandi, Christopher and Kroese, Dirk P and Johansen, Adam M},
  month = may,
  year = {2025},
  pages = {qkaf015},
}

@article{yang_surface_2024,
	title = {Surface phase diagrams from nested sampling},
	volume = {26},
	issn = {1463-9076},
	url = {https://doi.org/10.1039/d4cp00050a},
	doi = {10.1039/d4cp00050a},
	number = {18},
	urldate = {2026-10-13},
	journal = {Phys. Chem. Chem. Phys.},
	author = {Yang, Mingrui and Pártay, Livia B. and Wexler, Robert B.},
	month = may,
	year = {2024},
	pages = {13862--13874},
}

@misc{kaplan2025matpes,
  author        = {Kaplan, Aaron D. and Liu, Runze and Qi, Ji and Ko, Tsz Wai and
                   Deng, Bowen and Riebesell, Janosh and Ceder, Gerbrand and
                   Persson, Kristin A. and Ong, Shyue Ping},
  title         = {A Foundational Potential Energy Surface Dataset for Materials},
  year          = {2025},
  eprint        = {2503.04070},
  archivePrefix = {arXiv},
  doi           = {10.48550/arXiv.2503.04070},
  url           = {https://arxiv.org/abs/2503.04070},
}

@article{deng_chgnet_2023,
	title = {{CHGNet} as a pretrained universal neural network potential for charge-informed atomistic modelling},
	volume = {5},
	url = {https://www.nature.com/articles/s42256-023-00716-3},
	doi = {10.1038/s42256-023-00716-3},
	number = {9},
	journal = {Nat. Mach. Intell.},
	publisher = {Nature Publishing Group},
	author = {Deng, Bowen and Zhong, Peichen and Jun, KyuJung and Riebesell, Janosh and Han, Kevin and Bartel, Christopher J. and Ceder, Gerbrand},
	year = {2023},
	pages = {1031--1041},
}

@article{baldock_constant-pressure_2017,
	title = {Constant-pressure nested sampling with atomistic dynamics},
	volume = {96},
	url = {https://link.aps.org/doi/10.1103/PhysRevE.96.043311},
	doi = {10.1103/PhysRevE.96.043311},
	number = {4},
	journal = {Phys. Rev. E},
	author = {Baldock, Robert J. N. and Bernstein, Noam and Salerno, K. Michael and Pártay, Lívia B. and Csányi, Gábor},
	year = {2017},
	pages = {043311},
}

@article{martiniani_superposition_2014,
	title = {Superposition {Enhanced} {Nested} {Sampling}},
	volume = {4},
	url = {https://link.aps.org/doi/10.1103/PhysRevX.4.031034},
	doi = {10.1103/PhysRevX.4.031034},
	number = {3},
	urldate = {2022-02-06},
	journal = {Phys. Rev. X},
	author = {Martiniani, Stefano and Stevenson, Jacob D. and Wales, David J. and Frenkel, Daan},
	month = aug,
	year = {2014},
	pages = {031034},
}

@misc{wood_uma_2026,
	title = {{UMA}: {A} {Family} of {Universal} {Models} for {Atoms}},
	shorttitle = {{UMA}},
	url = {http://arxiv.org/abs/2506.23971},
	doi = {10.48550/arXiv.2506.23971},
	urldate = {2026-11-10},
	publisher = {arXiv},
	author = {Wood, Brandon M. and Dzamba, Misko and Fu, Xiang and Gao, Meng and Shuaibi, Muhammed and Barroso-Luque, Luis and Abdelmaqsoud, Kareem and Gharakhanyan, Vahe and Kitchin, John R. and Levine, Daniel S. and Michel, Kyle and Sriram, Anuroop and Cohen, Taco and Das, Abhishek and Rizvi, Ammar and Sahoo, Sushree Jagriti and Ulissi, Zachary W. and Zitnick, C. Lawrence},
	month = mar,
	year = {2026},
	note = {arXiv:2506.23971 [cs.LG]},
}

\end{document}